\documentclass[sn-mathphys,Numbered]{sn-jnl}

\usepackage{graphicx}%
\usepackage{multirow}%
\usepackage{amsmath,amssymb,amsfonts}%
\usepackage{amsthm}%
\usepackage{mathrsfs}%
\usepackage[title]{appendix}%
\usepackage{xcolor}%
\usepackage{textcomp}%
\usepackage{manyfoot}%
\usepackage{booktabs}%
\usepackage{algorithm}%
\usepackage{algorithmicx}%
\usepackage{algpseudocode}%
\usepackage{listings}%

\begin{document}

\title[DECICE]{DECICE: Device-Edge-Cloud Intelligent Collaboration Framework}

\author[1]{\fnm{Julian} \sur{Kunkel}}\email{julian.kunkel@gwdg.de}

\author[2]{\fnm{Christian} \sur{Boehme}}\email{christian.boehme@gwdg.de}

\author[1]{\fnm{Jonathan} \sur{Decker}}\email{jonathan.decker@uni-goettingen.de}

\author[3]{\fnm{Fabrizio} \sur{Magugliani}}\email{fabrizio.magugliani@e4company.com}

\author[4]{\fnm{Dirk} \sur{Pleiter}}\email{pleiter@kth.se}

\author[5]{\fnm{Bastian} \sur{Koller}}\email{koller@hlrs.de}

\author[6]{\fnm{Karthee} \sur{Sivalingam}}\email{karthee.sivalingam@huawei.com}

\author[7]{\fnm{Sabri} \sur{Pllana}}\email{Sabri.Pllana@forschungburgenland.at}

\author[8]{\fnm{Alexander} \sur{Nikolov}}\email{alexander.nikolov@synyo.com}

\author[9]{\fnm{Mujdat} \sur{Soyturk}}\email{mujdat.soyturk@marmara.edu.tr}

\author[10]{\fnm{Christian} \sur{Racca}}\email{christian.racca@top-ix.org}

\author[11]{\fnm{Andrea} \sur{Bartolini}}\email{a.bartolini@unibo.it}

\author[12]{\fnm{Adrian} \sur{Tate}}\email{adrian.tate@nag.co.uk}

\author[13]{\fnm{Berkay} \sur{Yaman}}\email{berkay.yaman@bigtri.net}


\affil[1]{\orgname{Universität Göttingen}, \city{Göttingen}, \country{Germany}}

\affil[2]{\orgname{GWDG}, \city{Göttingen}, \country{Germany}}

\affil[3]{\orgname{E4 Computer Engineering S.p.A.}, \city{Scandiano}, \country{Italy}}

\affil[4]{\orgname{KTH Royal Institute of Technology}, \city{Stockholm}, \country{Sweden}}

\affil[5]{\orgname{HLRS}, \city{Stuttgart}, \country{Germany}}

\affil[6]{\orgname{Huawei Technologies Düsseldorf GmbH}, \city{Düsseldorf}, \country{Germany}}

\affil[7]{\orgname{Forschung Burgenland GmbH}, \city{Eisenstadt}, \country{Austria}}

\affil[8]{\orgname{SYNYO GmbH}, \city{Wien}, \country{Austria}}

\affil[9]{\orgname{Marmara University}, \city{Istanbul}, \country{Turkey}}

\affil[10]{\orgname{Consorzio TOP-IX}, \city{Torino}, \country{Italy}}

\affil[11]{\orgname{Alma Mater Studiorum - Università di Bologna}, \city{Bologna}, \country{Italy}}

\affil[12]{\orgname{Numerical Algorithms Group Limited}, \city{Oxford}, \country{United Kingdom}}

\affil[13]{\orgname{BigTRI Bilisim A.S.}, \city{Istanbul}, \country{Turkey}}

\abstract{
DECICE is a Horizon Europe project that is developing an AI-enabled open and portable management framework for automatic and adaptive optimization and deployment of applications in computing continuum encompassing from IoT sensors on the Edge to large-scale Cloud / HPC computing infrastructures. In this paper, we describe the DECICE framework and architecture. Furthermore, we highlight use-cases for framework evaluation: intelligent traffic intersection, magnetic resonance imaging, and emergency response. 
}


\maketitle

\section{Introduction}

In the past decade, cloud computing has revolutionized computing by providing services over the Internet that can scale elastically for user requirements. The economics of scale have resulted in powerful computing and storage resources being provided at reduced costs. Furthermore, fault tolerance mechanism including replication allow to significantly enhance availability of services deployed in the cloud. Moreover, various application areas including smart home, smart city, or industrial automation come with needs that require computing services to be deployed closer to the end-user. Edge computing complements cloud computing by placing compute resources at the edge of the network~\cite{doi:10.1016/j.future.2019.02.050}. It also allows for connecting Internet of things devices that utilize alternative wireless protocols such as LoRaWAN.

While cloud and edge computing do share some characteristics, there are also characteristics and features that make edge computing unique. Placing computational resources in the proximity of users or data sources reduces latency and data transport and can even prevent shipping of sensitive data. The trend of placing computing resources outside of the data center has resulted in the emergence of the term compute continuum (or cloud continuum)~\cite{doi:10.1109/ACCESS.2022.3229185}. Having a continuum of resources available from the network edge (and beyond) to cloud and HPC data centers has attracted increased interest, also in the European context~\cite{etp4hpc:2023}. The emergence of a compute continuum is of particular interest for workflows that extend beyond the data center and have to leverage services that provide access to distributed edge, cloud, and HPC resources across federated infrastructures.

Due to inter-dependencies between distributed components and the potentially high level of heterogeneity of the compute, storage, and network resources, the scheduling of these resources becomes a huge challenge \cite{Carrion:2022}. The problem of finding the optimal scheduling decisions may be NP-complete. Moreover, dynamic factors such as faults, changing network capabilities and system topologies require constant monitoring and automatic adjustments to the placement of processes and data such that the manual creation of good static assignments of processes to hardware becomes infeasible. To overcome these challenges, we believe that new approaches are necessary that include the utilization of AI methods to facilitate timely resource allocation decisions for complex systems with thousands of nodes.

In this paper, we describe DECICE project \cite{decice-cordis:2022} that is developing an AI-based, open, and portable cloud management framework for automatic and adaptive optimization and deployment of applications in a compute continuum. The framework relies on holistic monitoring to construct a digital twin of the system that reflects all components and characteristics of the system. This digital twin will supply schedulers with information such that scheduling decisions for dynamic load balancing and data placements can be reached. This will lead to improved throughput and reduced resource consumption.  The framework's prototype will be integrated into Kubernetes, which is the most popular solution for orchestrating containers in the cloud. Kubernetes itself has no built-in support for edge computing. For this, DECICE relies on an extension of Kubernetes that is called KubeEdge~\cite{xiongExtendCloudEdge2018} and allows for transparently integrating edge nodes in the main cluster. Furthermore, to support HPC-like batch job execution in Kubernetes~\cite{liu2022fine} the prototype will explore the utilization of technology such as Volcano~\cite{Volcano:2023}. The DECICE project is working towards the following:

\begin{itemize}
\item
  Development of a cloud management framework based on
  open source and standards, which can seamlessly connect and deploy
  applications to devices across HPC, cloud and edge continuums.
\item
  Providing system administration and DevOps tools to
  access, control and manage the continuous service environment that
  includes the deployment of network, computing, data infrastructures
  and services.
\item
  Creating a dynamic digital twin for heterogeneous infrastructure by supplying relevant data via monitoring systems such that it can enable simulations and AI-based forecasting. 
\item
  Enhance application performance, reliability,
  throughput, energy efficiency and cost by automating decisions for
  application scheduling with the help of AI models.
\end{itemize}

The rest of this paper is structured as follows.
Section \ref{sec:context} provides context information.
An overview of DECICE framework is provided in Section \ref{sec:framework}.
Section \ref{sec:architecture} describes the architecture.
The strategy for DECICE evaluation via use-cases is described in 
Section \ref{sec:evaluation}.
Section \ref{sec:conclusion} concludes the paper.

\section{Context}
\label{sec:context}

For the rise of cloud computing, the ability to provide virtualized resources
has been critical.
Beyond virtual machines, containers have become a widely used technologies with
Kubernetes being the de facto orchestration platform for managing execution of 
containers at scale.
Kubernetes is an open source container orchestration platform that is able to 
dynamically handle containerized workloads across thousands of nodes by assigning 
resources and virtualizing networks.
One of the major reasons for the widespread usage of Kubernetes is its adaptability,
which enables anyone to extend, replace and remove various parts of these systems.

Before the times of cloud computing and container orchestration solutions,
scaling resources available to a service required proprietary solutions
for integrating load balancing capabilities and adding additional servers.
Such approaches had various short-comings as it was hard to roll out application
updates or to dynamically react to sudden bursts in service requests.

The virtualization offered by containers simplifies orchestration systems
like Kubernetes the choice of the physical
machine for running the service as long as it provides enough resources.
This works well for data centers with homogeneous hardware and fast and stable networks.
This situation changes in an environment with different types of hardware
as it is typically the case when including edge devices.

Another challenge is that Kubernetes has been designed for cloud computing
data centers and not for distributed environments involving edge devices.
This challenge has been addressed by KubeEdge~\cite{xiongExtendCloudEdge2018},
which extends Kubernetes such that it is able to use edge devices.
The key features of KubeEdge are its lightweight Kubernetes installation for less
powerful edge devices and a pull-based network infrastructure.
Via this network infrastructure, edge devices can appear as regular nodes on 
Kubernetes.
A key feature of KubeEdge is an implementation of OSI network protocol layers that
enables connecting edge devices and cloud servers within one virtual network.

With edge devices being added, the underlying set of hardware resources becomes
heterogeneous and, therefore, scheduling of resources significantly more complex.
Scheduling refers here to algorithms that allow -- based on the current state
of available resource utilization -- to decide on where to place a new container.
To the best of our knowledge there are currently no solutions available to
address the resulting scheduling challenges and to, e.g., decide on whether
to execute containers on cloud or edge nodes.
For this, the specific hardware characteristics of the available edge nodes
as well as the current network connectivity need to be taken into account.

Container scheduling is an active research field with various techniques being
explored.
Ahmad et al.~\cite{ahmadContainerSchedulingTechniques2021} classifies scheduling
techniques into one of four categories: 
(1) Mathematical modeling techniques,
(2) heuristic techniques,
(3) meta-heuristics techniques and
(4) machine-learning techniques.

Mathematical techniques provide a mathematical description of an optimization problem
that must be solved taking a set of constraints under account.
Integer linear programming is a widely used technique.
A typical challenge with such techniques is that they become too complex for
large-scale systems and take too long to solve. Heuristic techniques are generally scalable and fast, however, there is no guarantee for good scheduling decisions.

Meta-heuristic and machine-learning techniques have generally become
generally popular for optimization problems related to parallel computing systems
(see \cite{Memeti:2018} for a recent survey).
Meta-heuristic techniques are based on search strategies for finding good
but not necessary optimal solutions for a given optimization problem.
Typically nature-inspired algorithms are used \cite{Ferrandi:2010}.

Machine-learning techniques train a model using data collected in a given context.
These models later allow to make scheduling decisions based on what
has been learned earlier, for instance by using the model for making predictions
about the number of needed containers \cite{doi:10.1109/JSAC.2019.2895473}.

For DECICE it is planned to use machine-learning techniques for anomaly detection.
Such techniques have already been successfully used for HPC systems
(see, e.g., \cite{doi:10.1109/TPDS.2021.3082802}).
Furthermore, such techniques have been used for reducing energy-to-solution
(see, e.g., \cite{Memeti:2021}).
Within DECICE extending such approaches to the compute continuum will be explored.

Recently, a very significant increase of popularity of the digital twin concept
can be observed.
It is based on the realization of one or more virtual entities that describe a
physical entity.
A key feature of digital twins is the twinning of the physical and virtual entities,
i.e., the implementation of physical-to-virtual twinning and virtual-to-physical
twinning processes \cite{doi:10.1016/j.cirpj.2020.02.002}.
To the best of our knowledge, the concept of digital twinning has not been
applied to the case of cloud-edge systems.

\section{Framework Overview}
\label{sec:framework}

DECICE aims to develop an AI-based, open, and portable cloud management framework for automatic and adaptive optimization and deployment of applications in a federated infrastructure, including computing from the very large (e.g., HPC systems) to the very small (e.g., IoT sensors connected on the edge).

Working at such vastly different scales requires an intelligent management plane with advanced capabilities that allow it to proactively adjust workloads within the system based on their needs, such as latency, compute power, and power consumption.
Therefore, we envision an AI-model, which can use a digital twin of the resources available, to make real-time scheduling decisions based on telemetry data from the
resources.

The DECICE framework will be able to dynamically balance different workloads, optimize the throughput and latency of the system resources (compute, storage, and network) regarding performance and energy efficiency and quickly adapt to changing conditions.
The framework also provides the necessary tools and interfaces for the administrators and deployment experts to interface with all the infrastructure components and control them to achieve the desired result.

The integration of the DECICE framework with orchestration systems will
be done through open APIs to make it portable, modular, and
extensible.
The DECICE framework will be co-designed and evaluated through concrete use cases.

\begin{figure}[htb]
     \centering
     \includegraphics[width=\columnwidth]{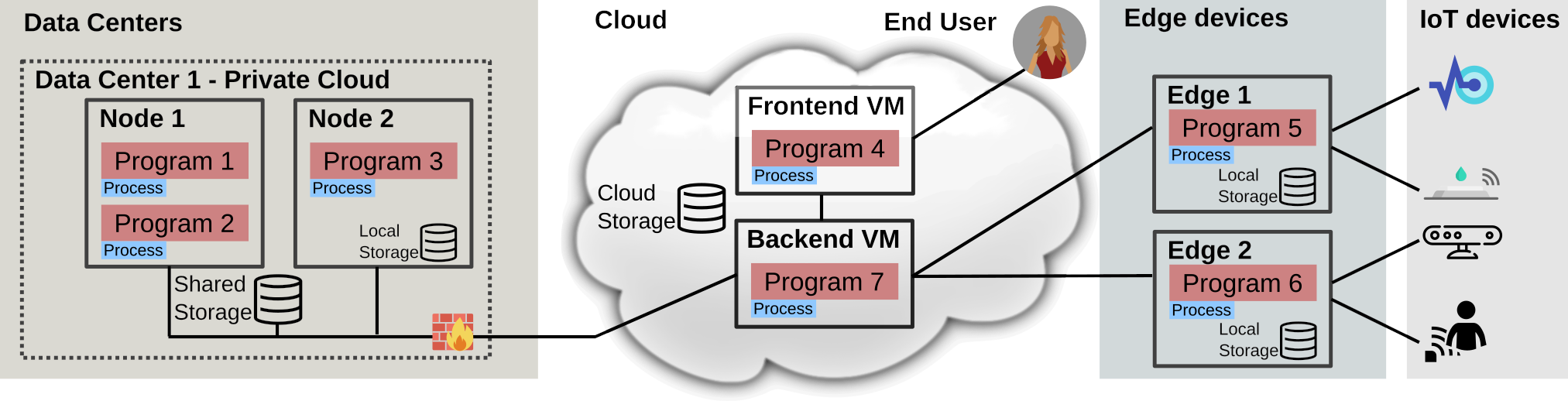}
     \caption{An example of computing continuum}
     \label{fig:topology}
\end{figure}

The system depicted in Figure \ref{fig:topology} can be seen as an example of a computing continuum executing a non-trivial application.
Consider an application that requires programs running on IoT devices to collect data, which is transferred via two edge nodes to the cloud nodes.
Due to heavy computation and privacy requirements, a data center might contribute compute resources and data volumes in a private cloud.
The results are then transferred back to the cloud where they are served to the end users. 
Finally, the cloud offers frontends that allow the end users to access the application and the results of the computation, for example, via a webpage.
The distribution of data across the continuum is another open challenge that is scheduled to tackle during the  performance test phase.

\section{DECICE Architecture}
\label{sec:architecture}

\begin{figure}[ht]
\centering
\includegraphics[width=\columnwidth]{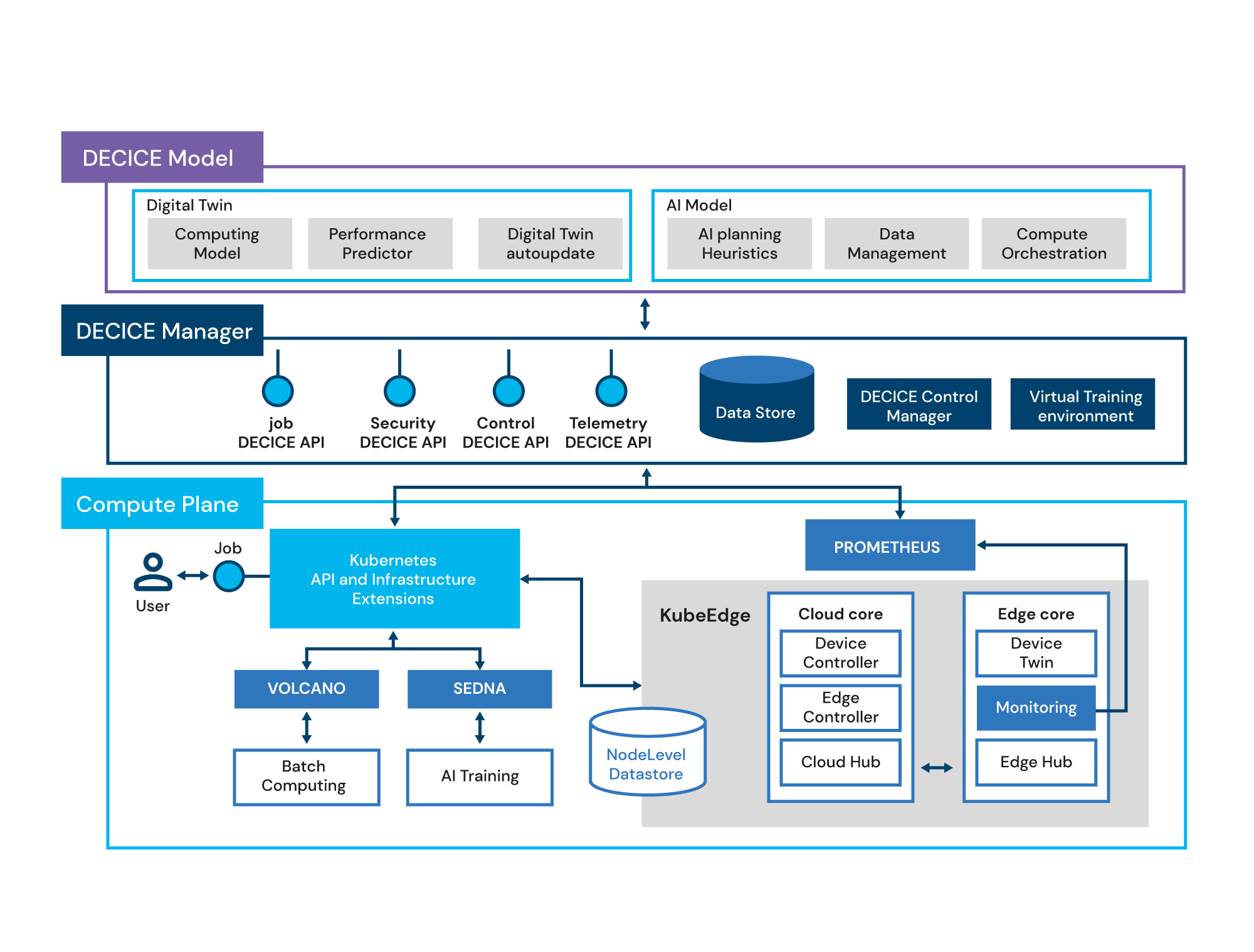}
\caption{High-level DECICE architecture}    
\label{fig:architecture}
\end{figure}

The DECICE framework is built on top of existing open-source
technologies that are already well-established in the community.

Kubernetes, KubeEdge, Prometheus, SEDNA and Volcano are part of the
\emph{Cloud Native Computing Foundation (CNCF)}, which is a hub for many open source projects from the cloud native ecosystem as well as over 250 members including many public cloud and enterprise cloud vendors.

Application users deploying to cloud-edge face multiple challenges and
are usually not experts in cloud management frameworks or heterogeneous
hardware resources. DECICE aims to help such users by minimizing the
effort required to efficiently utilize diverse hardware devices for
their application needs. For example, users of the DECICE framework will
be able to build hybrid edge/cloud/HPC workloads using standard containers,
knowing that the DECICE scheduler will allow them to specify data
locality, network latency or other constraints and that the AI-based
scheduler will place their jobs according to the requirements -- and move
them if needed. Primarily these capabilities will be provided via the
careful selection of base technologies and use of standardized
interfaces for interacting with DECICE. 
More importantly, via properly mapping and modeling the entire heterogeneous continuum into respective DECICE model.  

Users interact with DECICE via the standard Kubernetes interface, which
is extended with DECICE-specific extensions. This allows them to use an
existing, well-supported and familiar interface. Furthermore, it allows the
immediate movement of existing workloads into DECICE and minimizes the
``learning curve'' required for users to begin using DECICE.

The compute plane, the DECICE Manager and the DECICE Model as illustrated in Figure \ref{fig:architecture} will be
outlined in the following sections.

\subsection{Compute Plane}

The compute plane exists to run the user workloads. It is built on
existing open source technology, which is then integrated with the
DECICE components in the control plane. The primary component in the
compute plane is Kubernetes. It provides the basic
building blocks for running containerized jobs over multiple physical
systems. Kubernetes is further extended with KubeEdge to allow
the use and management of autonomous edge devices. Additionally,
Prometheus is used to gather metrics from devices via
Kubernetes/KubeEdge interfaces. Finally, SEDNA provides AI model
training capabilities, while Volcano adds the ability to run HPC
jobs.

\subsection{DECICE Manager}

The DECICE Manager integrates the Digital twin and AI models with the
orchestration system (e.g., Kubernetes) through its standard API. This
integration can be adapted to make the DECICE framework portable across
orchestration frameworks allowing integration with any commercial or open
source cloud solutions. Furthermore, the DECICE manager also includes a
data repository for storing metadata of infrastructure and applications
as well as monitoring data. This data is used to update the Digital
twin and train the AI-models. Moreover, the manager also provides a
synthetic training and test environment to simulate different
infrastructure and application deployment scenarios for the training of
AI-models.

\subsection{DECICE Model}

The Dynamic Digital Twin is a virtual representation of an actual or potential system at micro and macro level. It models the complete life-cycle of the system using simulation, real-time monitoring and enables forecasting, which is then used to make decisions. 
Creating a digital
twin of the heterogeneous cloud-edge infrastructure and the deployed
applications is one of the main goals of the project. The DECICE twin
models the compute, memory, storage and network resources as well as the
application tasks. Simulation and real time monitoring are used to
continuously update the model. AI models are then used to forecast the
system behavior in case of new deployment or a failure. AI modeling is
also used to continuously predict the application performance based on
the current state of the system and makes suggestions of optimization or
adaptation to increase performance, energy efficiency, reliability and
throughput. These suggestions are then acted upon by the DECICE
manager.

\subsection{Virtual Training Environment}

\begin{figure}[ht] 
    \centering
    \includegraphics[width=\textwidth]{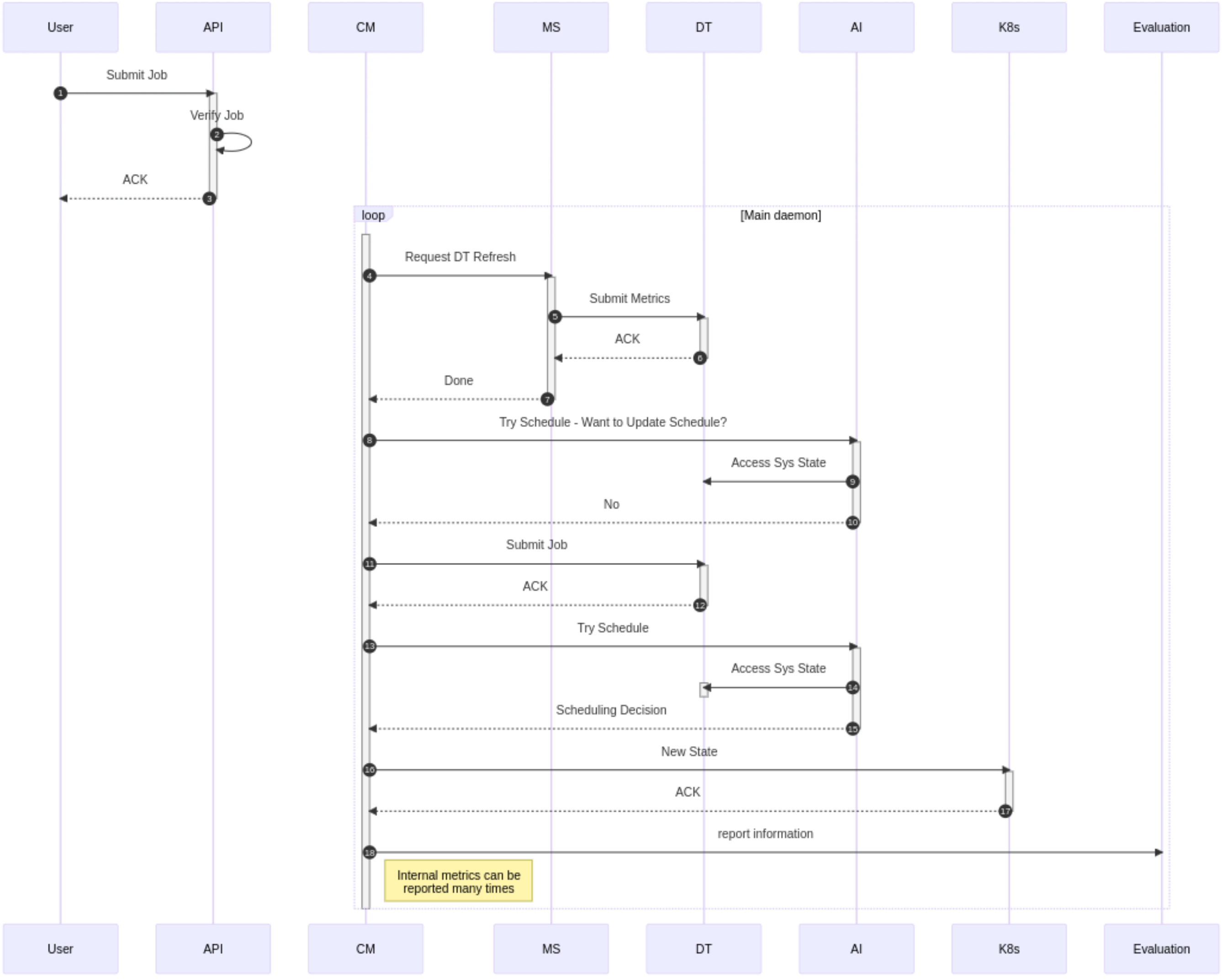}
    \caption{Sequence diagram: Virtualized Training; VTE: Virtual Training Environment; DT: Digital Twin}
    \label{sqd:virtual}
\end{figure}

In order to train, improve and test AI models for scheduling, we employ a Virtual Training Environment (VTE), which simulates a Digital Twin (DT) based on a pre-configured scenario.
The training process involves running the AI model under training against training scenarios, updating the scenario and evaluating the performance of the AI scheduler based on a set of metrics.
This process is depicted in Figure \ref{sqd:virtual} with the VTE loading a given training scenario, which in turn is used to configure the DT.
Within the main loop of the training, the VTE advances the scenario by progressing through the scenario, which in turn submits additional updates in the form of metrics and jobs to the DT.
Afterward, the scheduler accesses the DT to perform scheduling decisions.

As the VTE is responsible for advancing the scenario through the simulation and the AI scheduler itself has no concept of past states, it is possible to perform faster than real-time training.
This enables us to quickly train many instances of the AI scheduler in parallel and keep the best performing instances for further training.

The training loop used by the VTE in  Figure \ref{sqd:virtual} is similar to the scheduling loop that will be deployed on real systems. 
Instead of a VTE, the real system is controlled by the Cloud Manager (CM) and updates of the DT are pulled in from the Metrics System (MS) and the CM.
Furthermore, the scheduling decisions requested by the CM are then forwarded to the underlying Kubernetes (K8s) platform and realized.

Notably, the AI scheduler always views the entire DT, which also includes the entire job queue and state of all running jobs.
The AI scheduler has no obligation to immediately schedule a newly arrived job but its goal is to optimally utilize the underlying cluster for all active and queued jobs while honoring the user specifications.
This also includes rescheduling running jobs if a better placement becomes available by regularly reevaluating the current status of the system.

Overall, the AI scheduler is able to view and optimize the entire cluster through the DT.
A common heuristics-based scheduler operates by individually evaluating each queued job but could be expanded to view all jobs at once.
However, from our understanding, this would highly increase the complexity of the used heuristics without gaining value from a deeper understanding of the interplay of the components of the system as a machine learning-based scheduler would do.
We will also explore various deep learning strategies.

\subsection{CI/CD-Test Strategy}

The objective of our CI/CD systems is to verify that all components are functional individually and in composition.
Components developed in the context of the DECICE project, such as the Cloud Manager, the Digital Twin and the AI scheduler have their logic verified through individual component and mock testing.

As the DECICE systems integrate with an underlying orchestration platform that also evolves together with its ecosystem independently of DECICE, it becomes necessary to regularly perform integration tests.
For now, the prototype of the DECICE systems only targets Kubernetes as an orchestration platform.

By definition, the DECICE systems should be usable by administrators of heterogeneous compute systems that include cloud, edge and HPC resources.
Administrators may have additional requirements such as using a specific version of Kubernetes on their systems, which might be distinct from the version used to develop DECICE.
Therefore, the integration tests of DECICE cover a range of versions that are commonly used to ensure compatibility.

The testing of individual components is done via unit and mock testing.
Integration testing, however, requires spinning up a temporary Kubernetes cluster and deploying the components into it before running tests on them.
This allows using various configurations, versions and additional elements (e.g., edge and HPC nodes) over multiple test runs.

In practice, we are looking at GitLab runners for handling our CI/CD pipelines.
They can be integrated with Terraform and OpenStack to automatically roll out virtual machines, install Kubernetes on them and deploy our DECICE systems into it.
Such a system enables us to perform complete end-to-end tests of our software against multiple versions of Kubernetes, KubeEdge and Volcano.

\section{Evaluation Strategy}
\label{sec:evaluation}

\begin{figure}[ht]
\centering
\includegraphics[width=\columnwidth]{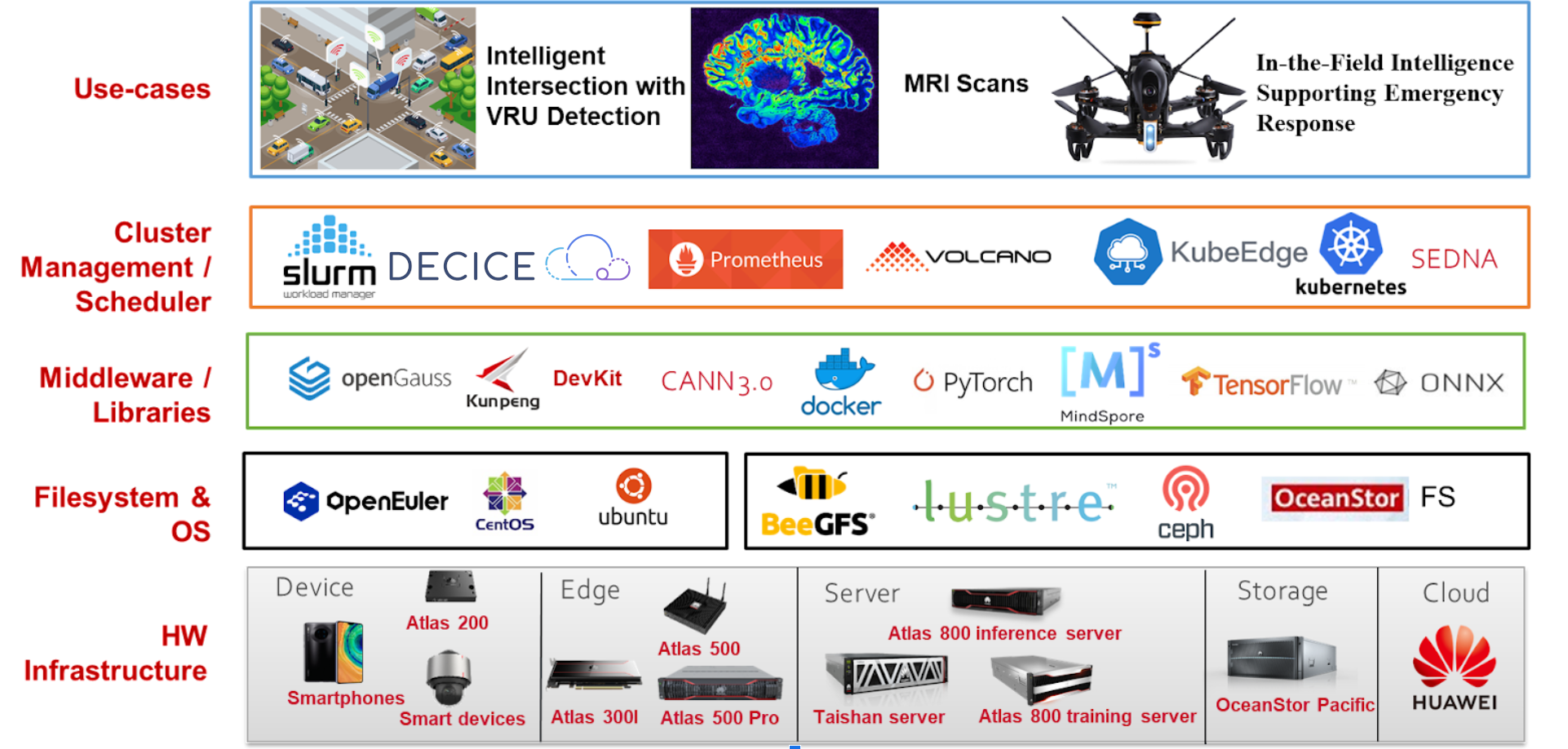}
\caption{Use-cases for DECICE evaluation: intelligent intersection, MRI, and emergency response}    
\label{fig:use-cases}
\end{figure}

In this section we describe our approach for evaluation of DECICE framework. The DECICE project adopts a co-design process in order to ensure that functionality and performance of the DECICE solutions do meet demands for real-life challenges. Figure \ref{fig:use-cases} depicts use-cases that are used for demonstrating the capabilities of technology developed in this project,

\begin{itemize}
    \item Intelligent intersection 
    \item Magnetic resonance imaging (MRI) 
    \item In-the-field intelligence supporting emergency response
\end{itemize}

\subsection{Intelligent intersection}

This use case focuses on real-time processing aspects in Connected \& Autonomous Driving and Cooperative Intelligent Transportation Systems (C-ITS). 
Connected and Autonomous Vehicles (CAVs) use their onboard sensors for object detection and situation awareness. Intersections introduce additional difficulties, challenges, and safety issues due to varying road users (cars, trucks, buses, pedestrians, bicycles, motorcycles, wheelchaired users, etc.) in varying numbers moving in different directions. 
Road user including Vulnerable Road User (VRU) detection is crucial and critical to ensuring safety and traffic efficiency. Road users which might be behind other objects or corners may not be detected by onboard sensors in vehicles. This poses risks to both road and driving safety and traffic efficiency. In addition to sharing the status information of the vehicles, it is aimed to share the information obtained from the sensors (e.g. camera) in the vehicles and at the intersections with the road users (vehicles, VRUs, etc.) in real-time. This approach is included by C2C-CC in Day 2 and Day3+ applications. 

Cameras positioned at an intersection continuously monitor the Region of Interest (RoI) and the captured images/videos are analyzed for object (vehicle, VRU, etc.) detection, which thereafter has to be shared with the road users (vehicles, VRUs). Processing of images and communication has to be in real-time (at low latencies) to provide safety and avoid possible accidents. 

The question arises on the level of hierarchy where the images will be processed. It could be done in the cloud, or edge, depending on the requirements and cost of doing that. Accuracy, latency, efficiency are the key technical performance metrics to measure. Data communication between the camera, and the central cloud introduces additional (communication) delay which is not acceptable. Therefore, there are general aims of carrying some fast processing, low latency, fast response applications at the edge, while learning models can work at the edge and cloud. On the other hand, resource utilization, cost, and traffic safety are the business-related performance metrics. DECICE provides an efficient, fast processing, low-latency, energy-efficient edge and cloud computing continuum that also considers the load at edge nodes due to the spatiotemporal variation of the users (vehicles, VRUs) and environmental conditions (light, illumination, weather conditions, etc). 

\subsection{Magnetic resonance imaging (MRI)}

MRI scans are a common tool for medical examination due to the noninvasive way to get a view of the inner body parts. Typically, MRI scans are conducted on edge nodes; they only cover individual body parts of a patient and create a high-resolution 3D-image of the body part using magnetic fields. Advances in technology lead to higher resolutions and, consequently, larger data sizes. Like other medical data, MRI scans often contain sensitive personal information about the patient. Therefore, special care with respect to storage, transfer and access is necessary. We propose the analysis of MRI scans as a use case for the DECICE framework to validate restrictions on storage and access rights for data. Furthermore, a fast and accurate analysis of the MRI scans is important, adding fast compute times to the list of requirements.

The analysis of the MRI scans poses difficulties, both from a compute and storage perspective. Each MRI scan consists of multiple slices of data, thus creating a complex object for analysis. Complex models for analysis can be challenging for the compute capabilities of edge devices. Thus, moving analysis jobs to the cloud or HPC system can be necessary. 
Managing access rights to the data, the storage location of data and preventing data loss are key components from the storage perspective.
Additionally, the network transfer can be a limiting factor creating a complex decision environment. 
Instead of having users manually figure out a solution, the DECICE schedule will map the data and tasks to suitable locations and also adapt to changes in system topology.

\subsection{In-the-field intelligence supporting emergency response}

In an emergency response, compute infrastructure and connectivity in the mission area can be seriously disturbed and unreliable.
To enable continuous monitoring of the affected areas and an effective response after a disaster struck, drones can be of help, data from multiple sensors in the drones and also additional data sources such as satellite images can be integrated to define mission areas or extend the area of interest. Quadcopters are an interesting opportunity but they have limited mission time. Also, image processing on board can be used but needs to be adapted to conditions, we need flexibility of computation. Recharging stations can be used to increase drones autonomy by also supporting upload of mission plans, data gathering, and communication. Missions can be made adaptive to weather. Coordination between operators and drone surveys to cover areas that are not covered by operators. 
The processing must be optimized, the integration of external sources such as satellite data to upgrade missions is possible. 

We aim at developing an open digital platform to support emergency response operators exploiting data from drones and satellites. 
The objective is to bring intelligence in the operational activities in the field, by providing the computation support for embedding ML algorithms to support drone autonomous flight as well as mission operations. 
To do so, the platform will implement an efficient and adaptive computation of AI algorithms in-the-field, orchestrating the processing and the communication between cloud/edge/drone in a fault-tolerant fashion, allowing the operator to control the data flows between the computation entities and providing the adaptivity in the system platform necessary to handle the emergency situation, such as size of the area to be explored, recognition tasks to be performed, communication link bandwidth, energy availability and weather conditions.

\section{Conclusions}
\label{sec:conclusion}

The cloud computing industry has grown massively over the last decade and with that new areas of application have arisen. Modern cloud applications have also become more complex as they usually run on a distributed computer system, split up into components that must run with high availability.

In the DECICE project, our goal is to develop an open and portable cloud-edge management framework for automatic and adaptive optimization of applications by mapping jobs to the most suitable resources in a heterogeneous system landscape and compute continuum. 
By utilizing holistic monitoring, we construct a digital twin of the system that reflects on the original system. An AI-scheduler makes decisions on placement of jobs and data as well as conducting job rescheduling to adjust to system changes dynamically. 
A virtual training environment is developed that generates test data for training of ML-models and the exploration of what-if scenarios. The portable framework is integrated into the Kubernetes ecosystem, co-designed and evaluated using relevant use cases on real-world heterogeneous systems.

\section*{Acknowledgments}
{This research is performed in the context of DECICE project that is funded by the European Union (grant agreement no. 101092582).}


\end{document}